\documentclass[prd,showpacs,amsmath,amssymb,11pt]{revtex4}
\usepackage{latexsym}
\usepackage{bm}

\begin{document}

\title{Massive Higher Derivative Gravity in $D$-dimensional Anti-de Sitter Spacetimes }

\author{\.Ibrahim G\"{u}ll\"{u} }
\email{e075555@metu.edu.tr}
\affiliation{Department of Physics,\\
             Middle East Technical University, 06531, Ankara, Turkey}

\author{Bayram Tekin}
\email{btekin@metu.edu.tr}
\affiliation{Department of Physics,\\
             Middle East Technical University, 06531, Ankara, Turkey}

\date{\today}

\begin{abstract}

We find the propagator and calculate the tree level scattering amplitude between two covariantly conserved 
sources in an Anti-de Sitter background for the most general $D$-dimensional quadratic, four-derivative, gravity 
with a Pauli-Fierz mass. We also calculate the Newtonian potential for various limits of the theory in flat space. 
We show how the recently introduced three dimensional New Massive Gravity 
is uniquely singled out among higher derivative models as a (tree level) unitary model and that its 
Newtonian limit is equivalent to that of the usual massive gravity in flat space.

\end{abstract}


\maketitle
            
\section{Introduction}

In gravity, there seems to be an insurmountable difficulty in reconciling 
renormalizability with unitarity in generic dimensions. By adding higher derivative terms, 
$\alpha R^2 +\beta R_{\mu \nu}^2$, to the four dimensional Einstein-Hilbert action, 
one gains renormalizability yet looses unitarity due to a non-decoupling ghost introduced by the $\beta$ term, 
without which one does not have a perturbatively renormalizable theory \cite{stelle}. 
The effect of non-unitarity in the Newtonian limit shows itself 
as a repulsive component to gravitational force between static sources. Because of this repulsive component, 
the theory has a better UV behavior. This is what usually happens in field theory: ghosts 
are introduced to make the theory better behaved, at least during the process of 
renormalization, yet, in a unitary theory they decouple at the end. 
Of course, bartering unitarity with renormalizability, as in the case of four dimensions, 
cannot be accepted. 

In three dimensions the situation seems to be better as was recently shown 
in \cite{townsend1,townsend2}: with the choice $ 8 \alpha + 3\beta = 0$ and a reversed sign 
Einstein-Hilbert term, one obtains a perturbatively renormalizable, 'unitary' theory in flat 
space \cite{nakason,oda,nakasonoda,deser}. But, it is not clear at all if this particular 
ratio between $\alpha$ and $\beta$ will survive renormalization at a given loop level, even at one-loop. 
The unitarity beyond {\it tree level } has to be checked. 
What is also interesting is that the 
linearized version of the theory has in its spectrum a massive graviton with helicities $\pm 2$. 
This fact sheds light, albeit only in three dimensions, to an old problem of finding 
a non-linear extension to the Pauli-Fierz mass term.   In fact, a formal equivalence of the 
Einstein-Hilbert-Pauli-Fierz gravity and the linearized version of the  ''New massive Gravity (NMG)" 
was shown in \cite{townsend1}. [Note that, in three dimensions besides this parity-preserving spin-2 theory,
we have the old  parity violating Topologically Massive Gravity (TMG) \cite{djt1,djt2}.]
Even though there does not seem to be a problem with this formal equivalence, one still needs to be careful 
about its physical meaning. It is clear that even the linearized version of the NMG theory is background 
diffeomorphism invariant, but the Pauli-Fierz theory  is only invariant under the Killing symmetries of the 
spacetime (in particular, the 2+1 dimensional Minkowski space). Therefore, a better understanding of the symmetries is 
needed. A quite interesting approach was put forward in \cite{deser}, where, in the linearized version 
of the NMG theory without the Einstein-Hilbert term, 
Weyl invariance of the action was shown. Therefore, at least in the linearized level, introduction 
of the Einstein-Hilbert term breaks this invariance and introduces 
a mass to the graviton. From this point of view, higher derivative terms provide the kinetic energy and 
the Einstein-Hilbert term  provides the mass  in this model, which also explains the bizarre 
sign change of the Einstein-Hilbert action. In retrospect, this is to be expected, 
pure Einstein's theory is non-dynamical and gives no propagation in three dimensions: 
at the linearized level, it is basically like the mass term in a scalar field theory 
which can only play role in the dynamics once a kinetic energy is introduced. Here the kinetic energy 
comes from the higher derivative terms.  So both in TMG and NMG, 
Einstein-Hilbert term gives rise to the mass in the linearized theories, by breaking  Weyl invariance 
( not the expected diffeomorphism invariance).  
This point of view could be important for  constructing  massive gravity theories in other dimensions. 

After \cite{townsend1}, several works appeared that were devoted to different aspects of this theory.
Especially, in AdS background that we shall be interested in, some classical solutions and conserved charges
were presented in \cite{clement,sun1, giribet, sun2, townsend2,oliva}. In this paper, we consider the most general quadratic model, 
augmented with a Pauli-Fierz mass term, in a $D$ dimensional (anti)-de Sitter background.    
We study the propagator structure and the tree level scattering  between two covariantly conserved sources in the theory 
obtained by linearizing the action
\begin{eqnarray}
I & = & \int d^{{D}}x\,\sqrt{-g}\left\{
\frac{1}{\kappa}R-\frac{2\Lambda_{0}}{\kappa}+\alpha R^{2}+\beta
R_{\mu\nu}^{^{2}}
+\gamma\left(R_{\mu\nu\sigma\rho}^{2}-4R_{\mu\nu}^{2}+R^{2}\right)\right\} \nonumber \\
 &  & +\int d^{D}x\,\sqrt{-g}\left \{-\frac{ M^{2}}{4\kappa } \left(h_{\mu\nu}^{2}-h^{2}\right)
+ {\cal {L}}_{\mbox{matter}}\right \},
\label{action}\end{eqnarray}
where $\Lambda_{0}$ is the bare cosmological constant and $\kappa$ is related to the 
$D-$dimensional Newton's constant and the $D-2$ dimensional solid angle by $\kappa \equiv 2\Omega_{D-2}G_D$. 
Including the number of dimensions, this 7-parameter theory is the most general, four derivative, quadratic model  
with various potentially interesting limits and discontinuities. [In the absence of the source terms and at the linearized level,
one can reduce the number of parameters in the action \cite{vas1,vas2}, but here for the sake of generality we shall work with 
(\ref{action}).] At this point, we assume nothing about the 
signs of the parameters in the action, moreover we will also allow them to vanish. Constraints will come from
the requirement of the tree-level unitarity and the non-existence of ghosts and tachyons.
For some specific dimensions,  certain terms will not contribute to the equations, 
for example, the $\gamma$ term  (Gauss-Bonnet  combination) is a total divergence in $D=4$ 
and vanishes identically for $D=3$, therefore, in three dimensions, the Riemann tensor carries no more information than 
the Ricci tensor. Also in  $D=3$, one can add the Chern-Simons term $\mu 
(\Gamma \partial \Gamma +\frac{2}{3} \Gamma^3)$ \cite{djt1,djt2} to extend our model, but here 
we will stick to (\ref{action}), since TMG is special to three dimensions. 
[See \cite{dest,tekinsari} for this case, without the higher curvature terms.] 
For $D=2$, the theory reduces to an $R^2$ model with a Pauli-Fierz term. Here, we will consider    
$D \ge 3$. Apart from the Pauli-Fierz mass term, the theory has general covariance. 

The spin-2 model defined by the linearization of (\ref{action}) is highly 
non-trivial. Needless to say, various limits  
have been studied  in the literature. Yet, there still appears 
interesting new models in certain limits of the above action. 
NMG, for $D=3$, in the case of a flat background,
and for ($M^2 =0$, $\gamma=0$) and  $8 \alpha +3\beta=0$, being one such example. 
One of our tasks in this paper is to explore in detail the full 7-parameter theory with particular care on the 
various discontinuities that appear in changing the order of limits when some of the parameters 
approach zero. One very well-known discontinuity is the so called van Dam-Veltman-Zakharov (vDVZ)
discontinuity: the fact that Einstein-Hilbert gravity in flat space is isolated from massive gravity in a
discontinuous manner, that is $M^2 \rightarrow 0$, at the tree level does not yield the correct General Relativity (GR) results.
But once a cosmological constant is introduced and  $M^2/\Lambda \rightarrow 0$ limit is taken, 
GR result is recovered \cite{higuchi,porrati,kogan,deserwaldron2,vainshtein}. [Another resolution of the discontinuity may follow even in 
flat space if the Schwarzschild radius of the 
scattering objects is taking as a second mass scale in the theory \cite{vainshtein}.] But, these are all at tree level, once quantum 
corrections are taken into account, discontinuity reappears \cite{duff}. Another related problem is the Boulware-Deser instability:
at the non-linear level, a ghost arises in massive gravity \cite{bouldeser}.

The lay out of the paper is as follows: In section II, we write down the linearized equations about an AdS  background and
discuss certain special limits, such as the partially massless case. In section III, we find the tree-level scattering amplitude 
and compute the Newtonian potential. 

\section{Linearized Equations}
The field equations that follow from (\ref{action}) are
\begin{eqnarray}
\frac{1}{\kappa}\left(R_{\mu\nu}-\frac{1}{2}g_{\mu\nu}R + \Lambda_0 g_{\mu\nu} \right )
+2\alpha R\left(R_{\mu\nu} -\frac{1}{4}g_{\mu\nu}R\right)+\left(2\alpha+\beta\right)\left(g_{\mu\nu}\square-\nabla_{\mu}\nabla_{\nu}\right)R\nonumber\\
+2\gamma\left[RR_{\mu\nu}-2R_{\mu\sigma\nu\rho}R^{\sigma\rho}+R_{\mu\sigma\rho\tau}R_{\nu}^{\;\;\sigma\rho\tau}
-2R_{\mu\sigma}R_{\nu}^{\;\;\sigma}-\frac{1}{4}g_{\mu\nu}\left(R_{\tau\lambda\sigma\rho}^{2}-4R_{\sigma\rho}^{2}
+R^{2}\right)\right]\nonumber\\+\beta\square\left(R_{\mu\nu}-
\frac{1}{2}g_{\mu\nu}R\right)+2\beta\left(R_{\mu\sigma\nu\rho}
-\frac{1}{4}g_{\mu\nu}R_{\sigma\rho}\right)R^{\sigma\rho} +\frac{M^{2}}{2\kappa }\left(h_{\mu\nu}-\bar{g}_{\mu\nu}h\right) =\tau_{\mu\nu}.
\label{fieldequations}
\end{eqnarray}
In the absence of the source, the `vacuum` 
( background $\bar{g}_{\mu \nu}$) is a non-singular solution to the field equations 
everywhere. This is the maximally symmetric (anti)-de Sitter Space 
with the Riemann, Ricci tensors and scalar curvature given respectively as
\begin{eqnarray}
\bar{R}_{\mu\rho\nu\sigma} = 
\frac{2\Lambda}{(D-1)(D-2)}\left(\bar{g}_{\mu\nu}\bar{g}_{\rho\sigma}
-\bar{g}_{\mu\sigma}\bar{g}_{\nu\rho}\right), \hskip 0.5 cm
\bar{R}_{\mu\nu} =\frac{2\Lambda }{D-2}\bar{g}_{\mu\nu}, \hskip 0.5 cm
\bar{R}  = \frac{2D \Lambda }{D-2}.
\label{background}  
\end{eqnarray}
Our discussion and notations follow \cite{dt1,dt2}.
All the contractions will be made with  
$\bar{g}_{\mu\nu}$ that has the signature $(-,+,...,+)$. Our conventions are 
$[\nabla_\mu,\nabla_\nu]V_\lambda=R_{\mu\nu\lambda}\,^\sigma
V_\sigma$, $R_{\mu\nu}=R^{\lambda}\,_{\mu\lambda\nu}$.
With these, one can find  $\Lambda $ in terms of $\alpha,\beta,\gamma,\kappa$
and $\Lambda_{0}$ (neglecting $M^2$), where $\Lambda $ is the effective
cosmological constant. From (\ref{fieldequations}) and  (\ref{background}) 
we have, 
\begin{equation}
\frac{\Lambda- \Lambda_0}{2 \kappa}+ \frac{D-4}{D-2}\left [ \frac{(D \alpha+\beta)}{D-2}
+\gamma \frac{D-3}{D-1}\right ]\Lambda^2
=0. 
\label{effectivecosmo}
\end{equation}
Generically, there are two solutions one of which vanishes in the absence of $\Lambda_0$, 
the other one nonvanishing even in this limit. But, lets first note the 
exceptions. For $D=4$, 
$\Lambda=\Lambda_{0}$.  Also, for $D=3$ if $3\alpha +\beta=0 $ we have 
$\Lambda =\Lambda_{0}$. Otherwise, in three dimensions 
$\Lambda_{1,2}=\frac{1\pm\sqrt{1-8\kappa\left(3\alpha+\beta\right)\Lambda_{0}}}
{4\kappa\left(3\alpha+\beta\right)}$ with a constraint
$1\geq 8 \kappa\left(3\alpha+\beta\right)\Lambda_{0}$. There is one more exceptional point: 
$\gamma=0$, $D\alpha +\beta=0$ theory also has $\Lambda=\Lambda_0$. 
[This is an interesting model whose action 
is given by the square of the traceless Ricci tensor. All the asymptotically AdS solutions 
( not just the globally AdS vacuum) 
have zero energy\cite{dt2}.] 
Finally, for the generic case we have
\begin{equation}
\Lambda=-\frac{1}{4\kappa f}\left[1\pm\sqrt{1+8\kappa
f \Lambda_{0}}\right],
\label{cosmogen}
\end{equation}
where $f \equiv \left(\alpha
D+\beta\right)\frac{\left(D-4\right)}{\left(D-2\right)^{2}}
+\gamma\frac{\left(D-3\right)\left(D-4\right)}{\left(D-1\right)\left(D-2\right)}$, for which the 
bound becomes as $8\kappa\Lambda_{0}f\geq-1$. 

Linearization of (\ref{fieldequations}) around the background metric, $g_{\mu\nu}=\bar{g}_{\mu\nu}+h_{\mu\nu}$, 
after using (\ref{cosmogen}), gives \cite{dt2}
\begin{eqnarray}
T_{\mu\nu}\left(h\right) & = & a {\mathcal{G}}_{\mu\nu}^{L}
  +  \left(2\alpha+\beta\right)\left(\bar{g}_{\mu\nu}\bar{\square}-\bar{\nabla}_{\mu}\bar{\nabla}_{\nu}
 +\frac{2\Lambda }{D-2}\bar{g}_{\mu\nu}\right)R^{L}\nonumber \\
 & + & \beta\left(\bar{\square}{\mathcal{G}}_{\mu\nu}^{L}-\frac{2\Lambda }{D-1}\bar{g}_{\mu\nu}R^{L}\right)
+ \frac{M^2}{2\kappa }\left(h_{\mu\nu}-\bar{g}_{\mu\nu}h\right),\label{linearizedfirst}
\end{eqnarray}
where we have defined
\begin{equation}
a \equiv \frac{1}{\kappa}+\frac{4\Lambda
D}{ D-2 }\alpha+\frac{4\Lambda}{D-1 }\beta+\frac{4\Lambda(D-3)(D-4)}{(D-1)(D-2)}\gamma .
\end{equation}
Here $T_{\mu\nu}\left(h\right)$ contains all the higher order terms as well as the source $\tau_{\mu \nu}$. 
$\mathcal{{G}_{\mu\nu}}^{L}$ is the linearization of the Einstein tensor (with the cosmological constant)
\begin{equation}
{\mathcal{G}}_{\mu\nu}^{L}=R_{\mu\nu}^{L}-\frac{1}{2}\bar{g}_{\mu\nu}R^{L}-\frac{2\Lambda }{D-2}h_{\mu\nu},
\label{einstein}\end{equation}
where the linearized Ricci tensor and the scalar curvature, $R^{L}=\left(g^{\mu\nu}R_{\mu\nu}\right)^{L}$, read 
\begin{equation}
R_{\mu\nu}^{L}=\frac{1}{2}\left(\bar{\nabla}^{\sigma}\bar{\nabla}_{\mu}h_{\nu\sigma}
+\bar{\nabla}^{\sigma}\bar{\nabla}_{\nu}h_{\mu\sigma}-\bar{\square}h_{\mu\nu}-\bar{\nabla}_{\mu}\bar{\nabla}_{\nu}h\right), \,\,
R^{L}=-\bar{\square}h+\bar{\nabla}^{\sigma}\bar{\nabla}^{\mu}h_{\sigma\mu}-\frac{2\Lambda}{ D-2 }h.
\label{ricci}\end{equation}
We also need the trace of (\ref{linearizedfirst}) which reads
\begin{equation}
\left [ \left( 4 \alpha (D-1) + D\beta \right)\bar{\square} - (D-2)\left (\frac{1}{\kappa} + 4 f \Lambda \right ) \right ] R^L 
- \frac{M^2}{\kappa }(D-1)h = 2 T . 
\label{tracedenk}
\end{equation}
It is clear that something special happens for $4 \alpha (D-1) + D\beta $ (that is the NMG point in $D=3$ ) 
but, before, we discuss this in detail by computing the tree level scattering between two sources, let us 
consider various limits of the theory at the linearized level, without the sources. 
Unitarity regions cannot be captured this way, but we can see the parameter ranges that rule out the tachyons.
At this point, the discussion depends on whether $M^2$ vanishes or not. Let us consider 
these cases separately. 

{\underline {\bf $M^2 \ne 0$ case } }

Divergence and the double divergence of  (\ref{linearizedfirst}) give
\begin{equation}
\bar{\nabla}^{\mu}h_{\mu\nu} - \bar{\nabla}_{\nu}h =0, \hskip 1 cm  
\bar{\nabla}^{\mu}\bar{\nabla}^{\nu}h_{\mu\nu}-\bar{\square}h = 0,
\label{divdenk}
\end{equation}
which lead to $ R^L= -\frac{2\Lambda}{ D-2 }h $. The discussion again bifurcates according $\Lambda =0$ or not.  
First consider the flat space case for which we have  $R^L=0$ and (\ref{tracedenk}) and the first equation of (\ref{divdenk}) force  
the field to be traceless and transverse.  The field equation for the remaining 
$(D+1)(D-2)/2$ independent components reduces to
\begin{equation}
\left ( \beta \partial^4 + \frac{1}{\kappa}\partial^2 -\frac{M^2}{\kappa } \right ) h_{\mu \nu}= 0,
\end{equation}
that describes two massive excitations with masses 
\begin{equation}
m_{\pm}^2= -\frac{1}{2\kappa \beta} \pm \frac{1}{2|\kappa \beta|}\sqrt{ 1+ 4 \beta M^2 \kappa },  
\label{2kutle}
\end{equation}
which are non-tachyonic if the parameters are properly picked. But, as we will see, this model is nonunitary.

Now assume that $\Lambda \ne 0$, then the trace equation gives
\begin{equation}
\left [ ( 4 \alpha (D-1) + D\beta ) \square - (D-2) \left ( \frac{1}{\kappa}+ 4 \Lambda f \right ) + \frac{M^2}{2 \kappa \Lambda } (D-1)(D-2) \right ] h = 0.
\end{equation}
which says that generically $h$ is a dynamical scalar field, unless the coefficient  of the D'Lambertian vanishes. Suppose, we pick up that special 
case, then, there are still two options: either $h=0$ or we have the partially massless point, that arises only in curved backgrounds for which a higher 
derivative gauge invariance appears \cite{deserwaldron} and the field has one less degree of freedom (DOF) compared to the massive one. The mass should be
tuned as 
\begin{equation}
M^2 = \frac{2 \Lambda \kappa }{D-1} \left ( \frac{1}{\kappa} - \frac{\Lambda \beta(D-4)}{D-1} + 4\Lambda\gamma\frac{(D-3)(D-4)}{(D-1)(D-2)} \right ),
\end{equation}
which is allowed to be negative in AdS as long as it satisfies the Breitenlohner-Freedman type bound. Apart from four dimensions, higher derivative terms
play role on the partially massless theory. 

{\underline {\bf $M^2 = 0$ case }}

The theory is now invariant under background diffeomorphisms $\delta_\xi h_{\mu \nu} = \bar \nabla_{\mu} \xi_{\nu} + \bar \nabla_{\nu} \xi_{\mu}$, since
$\delta_{\xi}{\mathcal{G}}_{\mu\nu}^{L}= 0$ and $\delta_{\xi} R^L=0$. Therefore, divergence and the double divergence 
do not give any constraint on $h_{\mu \nu}$. In this case, for $T =0$, (\ref{tracedenk}) gives dynamics to $R^L$ unless, the coefficient of the box term 
vanishes. In that special case, generically $R^L=0$, but it is clear that cosmological constant introduces another possibility : 
If $\frac{1}{\kappa}+ 4 \Lambda f =0$, then $R^L$ need not vanish. But this point does not seem acceptable since a gauge invariant object is left 
undetermined by the field equations.

\section{Tree-level Amplitude}

From now on, we consider the full theory (\ref{linearizedfirst}) and find the tree level scattering amplitude between two covariantly conserved 
sources. First we need to express $h_{\mu \nu}$ in terms of $T_{\mu \nu}$. But, since  
not all components of the field are independent, we decompose it in such a way that the physical parts 
will be determined by the source. The usual choice is to define
\begin{equation}
h_{\mu\nu} \equiv h_{\mu\nu}^{TT}+\bar{\nabla}_{(\mu}V_{\nu )} +\bar{\nabla}_{\mu}\bar{\nabla}_{\nu}\phi+\bar{g}_{\mu\nu}\psi,
\label{decompose}
\end{equation}
where $h_{\mu\nu}^{TT}$ is the transverse and traceless part. Symmetrization (with a 1/2 factor ) is implied in the vector part
$V_\mu$ which is divergence free. $\phi$ and $\psi$ are scalar functions.
Taking the trace, divergence and double divergence of Eq. (\ref{decompose}) one obtains
\begin{equation}
h=\bar{\square}\phi+D\psi, \hskip 1 cm 
\bar{\square}h=\bar{\square}^{2}\phi+\frac{2\Lambda}{\left(D-2\right)}\bar{\square}\phi+\bar{\square}\psi,
\label{handboxh}
\end{equation}
where we used $\bar{\nabla}^{\nu}\bar{\nabla}^{\mu}h_{\mu\nu}=\bar{\square}h$, which is {\it not } a gauge condition but imposed on us 
as a result of the nonzero mass term. Then, hitting the first equation of (\ref{handboxh}) with a $\bar{\square}$, one can eliminate  $\bar{\square} \phi$ 
with the help of the second equation as 
\begin{equation}
\bar{\square}\phi=\frac{\left(D-1\right)\left(D-2\right)}{2\Lambda}\bar{\square} \psi,
\label{phidenk}
\end{equation}
which then yields
\begin{equation}
h=\left ( \frac{\left(D-1\right)\left(D-2\right)}{2\Lambda}\bar{\square}+ D \right ) \psi.
\label{kdenk}
\end{equation}
From (\ref{tracedenk}), it follows that $\psi$ is determined by the trace of the energy momentum tensor
\begin{eqnarray}
\psi  =  \left\{\frac{\Lambda}{\kappa}+ 4 \Lambda f -c \Lambda \bar{\square}-
 \frac{M^2}{2 \kappa }\left(D-1\right)\right\} ^{-1}\left(\frac{\left(D-1\right)\left(D-2\right)}{2\Lambda}\bar{\square} + D\right)^{-1}T,
 \label{phiileT}\end{eqnarray}
where $ c \equiv \frac{4 (D-1 )\alpha}{D-2 } +\frac{ D\beta}{D-2}$. To find the transverse traceless part of the field in terms of the source, Lichnerowicz operator,  
$\triangle_{L}^{(2)}$ acting on spin-2 symmetric tensors is quite useful:
\begin{equation}
\triangle_{L}^{(2)}h_{\mu\nu}=-\bar{\square}h_{\mu\nu}-2\bar{R}_{\mu\rho\nu\sigma}h^{\rho\sigma}+
2\bar{R}^\rho\,_{(\mu} h_{\nu )\rho}
\label{lich}
\end{equation}
Some properties of this operator that we need were collected in \cite{porrati}
\begin{eqnarray}
\triangle_{L}^{(2)}\nabla_{(\mu}V_{\nu)} & = &
\nabla_{(\mu}\triangle_{L}^{(1)}V_{\nu)}, \hskip 0.5 cm  \triangle_{L}^{(1)}V_{\mu}
=\left(-\square+\Lambda\right)V_{\mu}, \hskip 0.5 cm
\nabla^{\mu}\triangle_{L}^{(2)}h_{\mu\nu} =  \triangle_{L}^{(1)}\nabla^{\mu}h_{\mu\nu}, \nonumber \\ 
\triangle_{L}^{(2)}g_{\mu\nu}\phi & = & 
g_{\mu\nu}\triangle_{L}^{(0)}\phi,  \hskip 0.5 cm \triangle_{L}^{(0)}\phi
=-\square\phi, \hskip 0.5 cm
\nabla^{\mu}\triangle_{L}^{(1)}V_{\mu}  = 
\triangle_{L}^{(0)}\nabla^{\mu}V_{\mu}.
\label{lichpro}
\end{eqnarray}
Using these we have
\begin{equation}
{\mathcal{G}}_{\mu\nu}^{LTT}=\frac{1}{2}\triangle_{L}^{(2)}h_{\mu\nu}^{TT}
-\frac{2\Lambda}{\left(D-2\right)}h_{\mu\nu}^{TT},
\end{equation}
which then leads to the desired equation
\begin{eqnarray}
h_{\mu\nu}^{TT} =  2 \left\{ (\beta \bar{\square} + a )( \triangle_{L}^{(2)} - \frac{4\Lambda}{D-2})
+ \frac{M^2}{\kappa}  \right \}^{-1}T_{\mu\nu}^{TT}.
\label{tracelessh}
\end{eqnarray}
The transverse traceless part of the energy momentum tensor can be found as ( after using the fact that it is covariantly conserved )
\begin{eqnarray}
T_{\mu\nu}^{TT}=T_{\mu\nu}-\frac{\bar{g}_{\mu\nu}}{D-1}T&+&\frac{1}{ D-1 }\left(\bar{\nabla}_{\mu}\bar{\nabla}_{\nu}
+\frac{2\Lambda\bar{g}_{\mu\nu}}{\left(D-1\right)\left(D-2\right)}\right)\nonumber\\&\times&\left(\bar{\square}
+\frac{2\Lambda
D}{\left(D-1\right)\left(D-2\right)}\right)^{-1}T. \label{denkTT}\end{eqnarray}
Finally using (\ref{phiileT}, \ref{tracelessh}, \ref{denkTT}), we can write the tree level scattering amplitude between two conserved sources
\begin{equation}
A=\frac{1}{4}\int
d^{D}x\:\sqrt{-\bar{g}}T'_{\mu\nu}\left(x\right)h^{\mu\nu}\left(x\right)=  \frac{1}{4} \int
d^{D}x\:\sqrt{-\bar{g}} \left ( T_{\mu\nu}'h^{TT\mu\nu}
+T'\psi \right ). \label{amplitude}
\end{equation}
For the sake of notational simplicity, lets suppress the integral for now: 
\begin{eqnarray}
4A & = & 2 T'_{\mu\nu}\left\{ (\beta \bar{\square} + a )( \triangle_{L}^{(2)} - \frac{4\Lambda}{D-2})
+ \frac{M^2}{\kappa}  \right \}^{-1} T^{\mu\nu} \nonumber \\  
&+& \frac{2}{D-1} T' \left\{ (\beta \bar{\square} + a)( \bar{\square} +\frac{4\Lambda}{D-2} ) - \frac{M^2}{\kappa} \right \}^{-1} T \nonumber \\
&-&\frac{4\Lambda}{(D-2)(D-1)^2} T'\left\{ (\beta \bar{\square} + a )( \bar{\square} +\frac{4\Lambda}{D-2} ) - \frac{M^2}{\kappa} \right \}^{-1} 
\left \{ \bar{\square} + \frac{2\Lambda D}{(D-2)(D-1)}\right \}^{-1} T  \nonumber \\  
&+&\frac{2}{(D-2)(D-1)} T'\left\{ \frac{1}{\kappa} + 4 \Lambda f - c \bar{\square}
- \frac{M^2}{2 \kappa \Lambda}(D-1) \right \}^{-1} \left\{ \bar{\square} + \frac{2\Lambda D}{(D-2)(D-1)} \right \}^{-1} T. 
\label{mainresult}
\end{eqnarray}
This is our main result from which we will consider various limits. For nonzero cosmological constant, this is quite a 
nontrivial integral. But we can figure out the particle spectrum of the theory by looking at the pole 
structure of the amplitude. Generically there are 4 poles which read as 
\begin{eqnarray}
\bar{\square}_1 & = & -\frac{2\Lambda D}{\left(D-1\right)\left(D-2\right)},\\
\bar{\square}_{2,3} & = &
\frac{1}{\beta} \left \{-\left(\frac{a}{2}+\frac{2\Lambda \beta }{\left(D-2\right)}\right)\pm\sqrt{\left(\frac{a}{2}
+\frac{2\Lambda \beta }{\left(D-2\right)}\right)^{2}- \beta\left(\frac{4\Lambda a}{\left(D-2\right)}
-\frac{M^2}{\kappa}\right)}\right \}, \\
\bar{\square}_4 & = & \frac{1}{c} \left ( \kappa^{-1} + 4 \Lambda f -\frac{M^2}{2 \kappa \Lambda}(D-1) \right ).
\label{poles}
\end{eqnarray}
Given these poles, finding the residues is easy but, in the most general form the expressions are rather cumbersome. 
The existence of the cosmological constant and the Pauli-Fierz mass term changes the picture drastically: Depending on the choice of the parameters, 
there could be tachyon and ghost-free models. Here, let us restrict our theory and discuss some interesting limits and compute the Newtonian 
potential between static sources for some of these limits. 

Looking at (\ref{mainresult}), it is clear that $M^2 \rightarrow 0$ and $\Lambda \rightarrow 0$ limits do not commute. 
In fact first taking the $\Lambda \rightarrow 0 $ limit, one encounters the vDVZ discontinuity. 
\begin{eqnarray}
4 A = -2T'_{\mu\nu}\left\{ \beta \partial^4 + \frac{1}{\kappa} \partial^2 -\frac{M^2}{\kappa}  \right\}^{-1}T^{\mu\nu}
+\frac{2}{D-1}T'\left\{\beta \partial^4 + \frac{1}{\kappa} \partial^2 -\frac{M^2}{\kappa}  \right\}^{-1}T,
\label{flatvdvz}
\end{eqnarray}
whose spectrum has two massive excitations with masses given by (\ref{2kutle}). To see its structure more explicitly,
one can rewrite it as 
\begin{eqnarray}
&&4 A =  \\
&-&\frac{2}{\beta (m_{-}^2-m_{+}^2 ) } \left \{ T'_{\mu\nu}\left ( \frac{1}{ \partial^2 -m_{+}^2}  - \frac{1}{ \partial^2 -m_{-}^2}  
\right ) T^{\mu\nu}
-\frac{1}{(D-1)}T'\left ( \frac{1}{ \partial^2 -m_{+}^2}  - \frac{1}{ \partial^2 -m_{-}^2 } \right )T \right \}. \nonumber 
\label{flatvdvz2}
\end{eqnarray}  
Unless $\beta=0$, we have a massive ghost. The Newtonian potential energy ($U$) between  $T'_{00}\equiv m_1\delta(x-x_1)$, $T^{00}\equiv
m_2\delta(x-x_2)$ in three and four dimensions can be obtained as
\begin{eqnarray}
U &=& \frac{1}{2\beta (m_{+}^2-m_{-}^2)}\frac{m_1 m_2 }{4\pi}[K_0(m_{-}r)-K_0(m_{+}r)]\hspace{2cm}D=3, \nonumber \\
U&=& \frac{m_1 m_2 }{3\beta (m_{+}^2-m_{-}^2) }\frac{1}{4\pi r}[e^{-m_{-} r}-e^{-m_{+}r }]\hspace{3.5 cm}D=4.
\end{eqnarray} 
where $r\equiv|\vec{x_1}-\vec{x_2}|$.  As $\beta \to 0$, the potential energies become
\begin{eqnarray}
U=-\frac{\kappa}{8\pi}m_1 m_2 K_0(M r)\hspace{2 cm} D=3,
\label{massive3d}
\end{eqnarray}
\begin{equation}
U= -\frac{4}{3}\frac {G m_1 m_2}{r}e^{-M r} \hspace{2 cm}D=4.
\end{equation}
The latter equation shows the famous discontinuity of massive gravity in flat space. [Note that we have used $\kappa =16 \pi G$, 
in four dimensions.] Let us stress that (\ref{massive3d}) is the Newtonian limit of massive gravity in three dimensions. 
It gives an attractive force as long $\kappa$ is positive. Unlike the four dimensional case, $M \to 0$ limit does not exist since, as 
$x \to 0$,  $K_0(x) \to -\ln(x/2) + \gamma_E $, which incidentally gives the expected $1/r$ force for small separation between the sources. 
So massive gravity gives the correct Newtonian limit in three dimensions where pure Einstein theory does not give any interaction.

Let us now consider the case which will lead us to the NMG model found in \cite{townsend1}. Take first $M^2=0$ then $\Lambda\to 0$.

\begin{eqnarray}
&&4A=-2 T'_{\mu\nu}\left \{\beta \partial^4 + \frac{1}{\kappa}\partial^2 \right \}^{-1}T^{\mu\nu}
+\frac{2}{(D-1)}T'\left \{\beta \partial^4 + \frac{1}{\kappa}\partial^2 \right \}^{-1} T\nonumber\\
&&-\frac{2}{(D-1)(D-2)}T'\left \{ c \partial^4 -\frac{1}{\kappa}\partial^2 \right \}^{-1} T
\label{ozel}
\end{eqnarray}
Generically there are three poles :
\begin{equation}
\partial_1^2 = 0, \hskip 1 cm \partial_2^2 = -\frac{1}{\kappa\beta} \hskip 1 cm, \partial_3^2 = \frac{1}{\kappa c}.
\end{equation}
For nonzero $\beta$, the requirement of unitarity singles out 
the $D=3$ and $8 \alpha + 3\beta =0$ theory in a highly nontrivial way. Let us explain how: not to have a tachyon, we should choose
$\kappa \beta < 0$. Looking at the residue of this pole, we see that we should have $\kappa < 0$ for unitarity. Computing the residue of the 
massless pole, we see that, with negative $\kappa$, $D=3$ must be chosen so that one does not have a massless ghost. Finally, the third pole 
is non-tachyonic if $c < 0$, but then, the residue of this pole requires $c > 0$ for unitarity. This is only possible if 
$c = 8 \alpha + 3 \beta =0$, which is the NMG theory. Newtonian limit of this model also reveals its rather unique structure. 
Let us for the moment compute the potential for generic $\alpha$ and $\beta$. 
\begin{eqnarray}
U=\frac{\kappa}{8\pi}m_1 m_2 \left ( K_0(m_g r)- K_0(m_0 r) \right ) \hspace{2cm} D=3,
\end{eqnarray}
where $m_g^2 \equiv -\frac{1}{\kappa \beta}$ and $m_0^2\equiv\frac{1}{\kappa(8\alpha+3\beta)}$. Clearly, $m_0$ 
is a massive ghost that gives a repulsive component. But, for NMG it decouples and one is left with an attractive force, 
since $\kappa <0 $. This result also confirms that, at this level, NGM has the same Newtonian limit as the usual massive gravity 
(\ref{massive3d}), if the Pauli-Fierz mass term is chosen as $M= m_g $. Beyond three dimensions, in flat space, massive ghost 
does not decouple unless $\beta =0$. As an example, let us look at $D=4$:
\begin{eqnarray}
U=-\frac{G m_1 m_2}{r} \left (1 -\frac{4}{3} e^{-m_g r}+\frac{1}{3}e^{-m_a r} \right ), 
\end{eqnarray}
where $m_a^2\equiv\frac{1}{2\kappa(3\alpha+\beta)}$. The middle, repulsive term signals the ghost problem 
\cite{stelle}.

\section{Conclusion}

We have studied the most general quadratic gravity with a Pauli-Fierz mass in $D$ dimensional (anti)-de Sitter space. 
From the tree level scattering amplitude that we found, one can study various limits. In flat space, we computed the Newtonian limits 
for various models including the New Massive Gravity that was recently introduced. Non-unitarity of the 
NMG theory comes in a highly non-trivial way and does not extend beyond three dimensions, in flat space. 
The cosmological constant changes the picture drastically, one needs to further study 
in detail the unitary regions. Especially in the NMG theory, as we mentioned in the introduction, unitarity beyond 
the tree level has to be checked, it is not clear at all if the condition $8\alpha + 3\beta =0$ will survive renormalization, 
nor it is clear that Boulware-Deser instability in the full non-linear theory is avoided.
We intend to address these problems in a future work.

\section{\label{ackno} Acknowledgments}
We would like to thank S. Deser for a useful discussion and for critical reading of the manuscript. 
This work is partially supported by  T{\"U}B\.{I}TAK Kariyer Grant 104T177.

\end{document}